\documentclass{article}
\usepackage[affil-it]{authblk}
\usepackage{natbib}
\usepackage{hyperref}
\usepackage{graphicx}
\usepackage{algorithm}
\usepackage[noend]{algpseudocode}
\usepackage{amsthm,amsmath,amssymb}
\usepackage{multirow}
\usepackage{subcaption}
\usepackage[top=1in, bottom=1in, left=1in, right=1in]{geometry}

\makeatletter
\def\BState{\State\hskip-\ALG@thistlm}
\makeatother

\newtheorem{them}{Theorem}

\author{Wei Jiang}
\author{Weichuan Yu
  \thanks{Mail: \texttt{eeyu@ust.hk}}}

\affil{Department of Electronic and Computer Engineering, The Hong Kong University of Science and Technology, Clear Water Bay, Kowloon, Hong Kong, China}

\date{}
\title{Controlling the joint local false discovery rate is more powerful than meta-analysis methods in joint analysis of summary statistics from multiple genome-wide association studies}
\begin{document}

\maketitle

\subsection*{\centering Abstract}
{\em
In genome-wide association studies (GWASs) of common diseases/traits, we often analyze multiple GWASs with the same phenotype together to discover associated genetic variants with higher power. Since it is difficult to access data with detailed individual measurements, summary-statistics-based meta-analysis methods have become popular to jointly analyze data sets from multiple GWASs.

In this paper, we propose a novel summary-statistics-based joint analysis method based on controlling the joint local false discovery rate (Jlfdr). We prove that our method is the most powerful summary-statistics-based joint analysis method when controlling the false discovery rate at a certain level. In particular, the Jlfdr-based method achieves higher power than commonly used meta-analysis methods when analyzing heterogeneous data sets from multiple GWASs. Simulation experiments demonstrate the superior power of our method over meta-analysis methods. Also, our method discovers more associations than meta-analysis methods from empirical data sets of four phenotypes. The R-package is available at: \url{http://bioinformatics.ust.hk/Jlfdr.html}.
}

\section{Introduction}
Understanding genetic mechanisms of common diseases and traits is important in biological and medical research. The goal of genome-wide association studies (GWASs) is to discover the susceptibility of single nucleotide polymorphisms (SNPs) to common diseases/traits \citep{altshuler2008genetic}. Due to decreasing genotyping costs \citep{perkel2008snp}, constantly emerging successful stories \citep{klein2005complement, kraft2010gwas} and efforts of the GWAS consortiums \citep{burton2007genome, schizophrenia2014biological}, more and more GWASs have been conducted for common phenotypes \citep{welter2014nhgri}.

Analyses of GWAS results show that the identified associations can only explain a small part of the additive genetic variances. This is referred to as the ``missing heritability'' problem \citep{manolio2009finding}. The hints of hidden heritability \citep{gibson2010hints, yang2010common} and the estimated distribution of common SNPs' effect sizes \citep{park2010estimation} suggest that common diseases/traits are influenced by thousands of SNPs with small effects. To discover these genetic variants with small effects, we need to improve studies' power. Jointly analyzing data sets from multiple GWASs on the same diseases in the same population provide an opportunity to improve the power.

There are two kinds of joint analysis methods: individual-level joint analysis and summary-statistics-based joint analysis. Individual-level joint analysis uses individual-level genotype data from all studies.  One such example is mega-analysis \citep{ripke2013mega}, which pools all data together. Summary-statistics-based joint analysis only uses summary statistics from different studies. Since individual-level genotype data is difficult to access, summary-statistics-based analysis is widely used in joint analysis. The most commonly used method of summary-statistics-based joint analysis is meta-analysis \citep{evangelou2013meta}, which derives a new statistic for each SNP using summary statistics from multiple studies.

\begin{figure}[!htbp]
    \centering
    \includegraphics[width=0.90\textwidth]{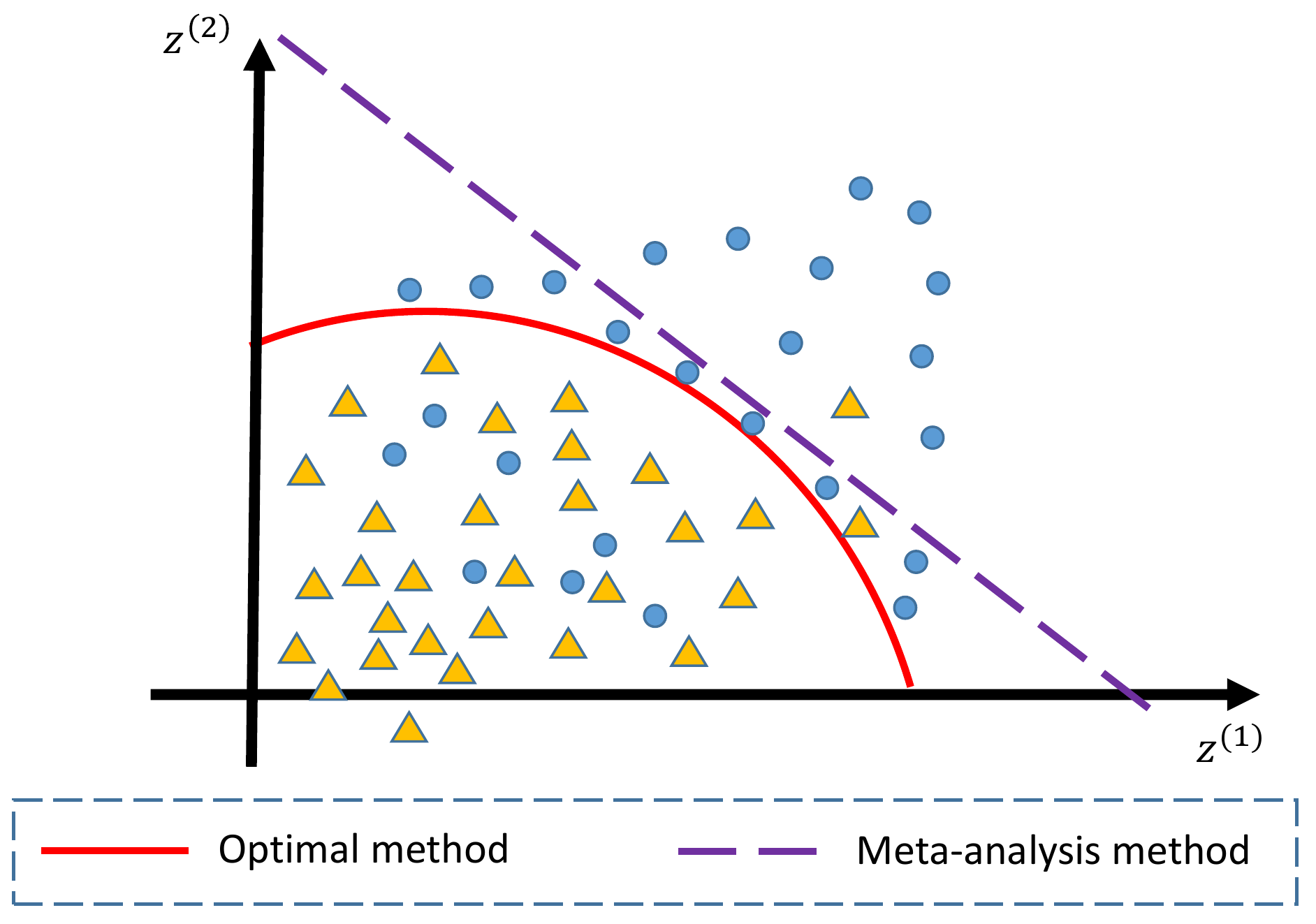}
    \caption{\textbf{Rejection boundaries determined by different summary-statistics-based joint analysis methods: the optimal method and the meta-analysis method.} Assume we jointly analyze data from two GWASs. For simplicity, we assume the tests are one-sided. We plot the test statistic pair $(z^{(1)}, z^{(2)})$ into the coordinate plane. A SNP at the upper right corner shows more significant association than a SNP at the bottom left corner. The true associated SNPs are plotted with blue circles, and the false associated SNPs are plotted with yellow triangles. For each rejection boundary, the SNPs in the upper right region are discovered. All three analysis methods have the same false discovery proportion (10\%). The optimal method has more empirical power (red solid line, 72\%) than the meta-analysis method (purple dashed line, 36\%).}\label{Figure1}
\end{figure}

Our focus in this paper is to study summary-statistics-based joint analysis methods. More specifically, we like to study which joint analysis method provides the highest power for a given false discovery rate level. Figure \ref{Figure1} illustrates our motivation.

Our major contribution in this paper is that we propose a novel summary-statistics-based joint analysis method based on controlling the joint local false discovery rate (Jlfdr). The Jlfdr generalizes the concept of the local false discovery rate \citep{efron2005local} from the analysis of single study to the joint analysis of multiple studies. We prove that our method is the most powerful summary-statistics-based joint analysis method for a given false discovery rate level. In particular, the Jlfdr-based method is more powerful than commonly used meta-analysis methods when analyzing heterogeneous data sets from multiple GWASs.

The rest of this paper is organized as follows. In Section \ref{methods}, we will first give the mathematical formulation of summary-statistics-based joint analysis methods. We will prove that the most powerful summary-statistics-based joint analysis method should control the Jlfdr. Then we will give implementation details of the Jlfdr-based method under the Gaussian mixture model. We will also discuss the relationship between the Jlfdr-based method and meta-analysis methods. In Section \ref{results}, we will use simulation experiments to demonstrate that Jlfdr-based method outperforms meta-analysis methods in terms of achieving higher power. Then we will show the empirical results using four different data sets.
In Section \ref{discussion}, we will discuss limitations of our current method. Section \ref{conclusion} concludes the paper.

\section{Method}\label{methods}
\subsection{Notations and criteria}
Our method deals with a multiple GWAS setting. For simplicity, we illustrate the concepts with a two-GWAS setting. We use parenthesized superscript ``$(j)$'' to denote the study index. For example, the sample sizes in study 1 and 2 are $n^{(1)}$ and $n^{(2)}$, respectively. We use subscript $i$ ($i=1,\dots, m$, $m$ is the total number of genotyped SNPs) to denote the SNP index.

To detect associations, we construct a null hypothesis for each SNP, in which association is assumed nonexistent. Assume we use a $z$-value scheme to detect associations between SNPs and the phenotype, i.e., the test statistics follow a standard normal distribution under a null hypothesis. We use $\widehat{\mu}^{(j)}$ to denote the observed effect size in study $j$. The asymptotically standard error of $\widehat{\mu}^{(j)}$ is $\sigma^{(j)}$. Correspondingly, the test statistic in study $j$ is $z^{(j)}=\widehat{\mu}^{(j)}/\sigma^{(j)}$. The underlying expected effect size is $\mu^{(j)}$. The expected effect size of the same SNP may vary in different studies due to heterogeneity. The test statistic $Z^{(j)}$ (uppercase letter indicates a random variable) follows an $N(\mu^{(j)}/\sigma^{(j)},1)$ distribution. We use $\pmb{z}$ to represent the vector of test statistics in all studies, i.e., $\pmb{z}=(z^{(1)},z^{(2)})^T$. Similarly, we use $\pmb{\mu}$ to represent the vector of expected effect sizes in all studies, i.e., $\pmb{\mu}=(\mu^{(1)}, \mu^{(2)})^T$.

We further assume $m_0$ SNPs have no association with the phenotype and $m_1$ SNPs have associations. Thus, the null proportion reads $\pi_0=m_0/m$ ($0\leq \pi_0\leq 1$). We use $\mathcal{H}_0$ and $\mathcal{H}_1$ to denote the null hypothesis and the alternative hypothesis, respectively.


In the joint analysis of summary statistics from multiple GWASs, we assume that $R$ of the $m$ hypotheses are rejected. There are $V$ false positives and $S$ true positives (i.e., $V+S=R$). Table \ref{contTable} summarizes the numbers of hypotheses in the different categories.

\begin{table}[!htbp]
\centering
\caption{\textbf{The status of all hypotheses in the joint analysis}. The letter in each cell denotes the count of the hypotheses in each category. }\label{contTable}
\begin{tabular}{c|cc|c}
& $\mathcal{H}_0$ is true & $\mathcal{H}_0$ is false & Total\\ \hline
$\mathcal{H}_0$ is rejected & $V$ & $S$ & $R$\\
$\mathcal{H}_0$ is not rejected & $U$ & $T$ & $m-R$\\ \hline
Total & $m_0$ & $m_1$ & $m$
\end{tabular}
\end{table}

When testing multiple hypotheses, it is very easy to have false positives by random chance. This problem is known as the ``multiplicity'' problem. Many criteria are proposed to address the multiplicity problem. We present an incomplete list of these criteria in Table \ref{criteria} (a). Let's define the false discovery proportion (Fdp) as $V/(R\vee 1)$ with ``$\vee$'' denoting the maximum operation. Fdp is an unknown quantity in real cases. The classical false discovery rate (FDR) is the expectation of the Fdp. Controlling the FDR is more powerful than controlling the family-wise error rate (FWER). The Bayesian false discovery rate (Fdr) is the expected value of the Fdp given $R>0$. Compared to FDR, Fdr is conditional on $R>0$ since we are only interested in controlling false positives when $R>0$. We adopt Fdr in this paper as the criterion to avoid a plethora of false positives.



\begin{table}[!htb]
    \caption{Different criteria for evaluating a rejection region in multiple testing scenario. Here $\mathcal{R}$ is the rejection region in the analysis. $\pmb{\mu}$ denotes effect sizes. $V$ and $R$ as well as other notations are explained in Table \ref{contTable}. ``$\vee$'' denotes the maximum operation.}\label{criteria}
    \begin{subtable}{\linewidth}
      \centering
        \caption{Different criteria for controlling false positives in multiple testing scenario.}
        \begin{tabular}{c|c|c}
            Criteria & Mathematical Definitions & References \\ \hline
            Family-wise error rate (FWER) & $\text{FWER}(\mathcal{R})=P(V\geq 1)$ & \citet{Tukey1953}\\
            False discovery rate (FDR) & $\text{FDR}(\mathcal{R})=E(V/(R\vee 1))$ & \citet{benjamini1995controlling}\\
            Bayesian false discovery rate (Fdr) & $\text{Fdr}(\mathcal{R})=E(V/R\big| R>0)$ & \citet{storey2003positive}\\  \hline
        \end{tabular}
    \end{subtable}%
    \\
    \begin{subtable}{\linewidth}
      \centering
        \caption{Different criteria for measuring the amount of true positives in multiple testing scenario.}
        \begin{tabular}{c|c|c}
            Criteria & Mathematical Definitions & References \\ \hline
            Power & $\beta(\mathcal{R},\pmb{\mu})=P(\pmb{z}\in \mathcal{R} \big| \mathcal{H}_1,\pmb{\mu})$ & \citet{Neyman1933}\\
            Bayesian power & $\eta(\mathcal{R})=P(\pmb{z}\in \mathcal{R} \big| \mathcal{H}_1)$ & \citet{kruschke2010believe}\\\hline
        \end{tabular}
    \end{subtable}
\end{table}

In addition to controlling false positives, we also need a criterion to measure the amount of true positives when evaluating a rejection region. A direct concept is power. The classical definition of power is a function of a given effect size as shown in the first row of Table \ref{criteria}(b). Since effect sizes of associated SNPs are different and unobserved, the actual power values are unknown. The Bayesian power removes the dependence of power on effect size by taking the expectation of the empirical power, which is defined as $S/m_1$ ($m_1>0$). We list the definitions of the power and the Bayesian power in Table \ref{criteria} (b). In this paper, we use the Bayesian power as the criterion to measure the amount of true positives.

Both Fdr and Bayesian power are functions of the rejection region $\mathcal{R}$. For two different rejection regions with the same Fdr level, we prefer the region with higher Bayesian power because it can find more true associations without increasing the proportion of false positives in the findings. Thus, we propose a joint analysis method determining the optimal rejection region when controlling the Fdr at a certain threshold $q$, i.e.
\begin{eqnarray}
\max_{\mathcal{R}}&\ & \eta(\mathcal{R})\nonumber\\
s.t. &\ & \text{Fdr}(\mathcal{R})\leq q.\label{Optimization}
\end{eqnarray}
Here $\eta(\mathcal{R})$ denotes the Bayesian power. Actually, when controlling the Fdr at the same threshold, meta-analysis methods can also be regarded as the solutions to the above optimization problem with further constraint about the form of $\eta(\mathcal{R})$, i.e.
\begin{eqnarray}
\max_{\mathcal{R}_C}&\ & \eta(\mathcal{R}_C)\nonumber\\
s.t. &\ & \text{Fdr}(\mathcal{R}_C)\leq q \nonumber \\
&\ & \mathcal{R}_C= \{\pmb{z}\big| |g(\pmb{\alpha},\pmb{z})|\geq C\}.\label{MetaOptimization}
\end{eqnarray}
Here $\pmb{\alpha}=(\sqrt{n^{(1)}},\sqrt{n^{(2)}})^T$, and $g$ is a function which has different forms in different meta-analysis methods. We will give the explicit forms of the function $g$ in meta-analysis methods in subsection \ref{relation}. Also, we will discuss the relationship between our proposed method and meta-analysis methods in detail in that subsection. In the next subsection, we will present the solution to the optimization problem in Eq. (\ref{Optimization}).

\subsection{Jlfdr and optimal rejection region}
To derive the solution to the optimization problem in Eq. (\ref{Optimization}), we need to introduce the concept of joint local false discovery rate (Jlfdr) first. Jlfdr is a simple extension of the local false discovery rate \citep{efron2005local} from the analysis of single study to the joint analysis of multiple studies. It reads as
\begin{equation}
\text{Jlfdr}(\mathbf{z})=P(\mathcal{H}_0 \big| \mathbf{z}),
\end{equation}
which is the posterior probability of a null hypothesis, given the observed summary statistic vector $\mathbf{z}$.

The relationship between Jlfdr and Fdr is (see the Supplementary Note for details)
\begin{equation}
\text{Fdr}(\mathcal{R})=E(\text{Jlfdr}(\mathbf{z})\big| \mathbf{z}\in \mathcal{R}).\label{rel}
\end{equation}
In other words, Fdr is the expectation of Jlfdr, given that the test statistic vector is in the rejection region $\mathcal{R}$.

Let us define a rejection region $\mathcal{R_O}=\{\mathbf{z} \big| \text{Jlfdr}(\mathbf{z})\leq t(q)\}$, where $t(q)$ is a threshold such that $\text{Fdr}(\mathcal{R_O})=q$. We have the following theorem:

\begin{them}\label{them1}
For any rejection region $\mathcal{R}$ with $\text{Fdr}(\mathcal{R})\leq q$, we have $\eta(\mathcal{R}) \leq \eta(\mathcal{R_O})$.
\end{them}

We show the proof of Theorem \ref{them1} in the Supplementary Note. Theorem \ref{them1} shows that $\mathcal{R_O}$ is the most powerful rejection region when controlling Fdr at $q$. This gives us a clue that we can improve the power of summary-statistics-based joint analysis by controlling the Jlfdr. In the next section, we shall provide details of the implementation of the Jlfdr-based method under the Gaussian mixture model.

\subsection{Implementation of Jlfdr-based method under the Gaussian mixture model}
The hints of hidden heritability \citep{gibson2010hints, yang2010common} and the estimated distribution of common SNPs' effect sizes \citep{park2010estimation} suggest that thousands of common SNPs with small effect sizes are associated with complex diseases. A natural prior to depict this ``infinitesimal model'' \citep{gibson2012rare} is Gaussian distribution with mean $0$ and variance $\sigma_0^2$. We assume the effect sizes of associated SNPs have this prior distribution. Since we don't know which SNP is associated with diseases, we propose the following two-component mixture model to describe the prior distribution of effect sizes:
\begin{equation}
\mu\sim \pi_0 \delta_0+(1-\pi_0) N(0, \sigma_0^2),
\end{equation}
where $\delta_0$ is the unit point mass distribution at zero.

There may be some heterogeneity in different studies. The effect sizes of the same SNP may vary in different studies. We assume the effect sizes of the same associated SNP in different studies are normally distributed with mean $\mu$ and variance $\tau \sigma_0^2$, i.e., $\mu^{(j)}\big| \mathcal{H}_1\sim N(\mu, \tau\sigma_0^2)$. The distribution of the effect size vector $\pmb{\mu}=(\mu^{(1)},\mu^{(2)})^T$ is
\begin{equation}
\pmb{\mu} \sim \pi_0 \pmb{\delta}_0+(1-\pi_0) N_2\left(\pmb{0}, \left( \begin{array}{cc}
(\tau+1)\sigma_0^2 & \sigma_0^2 \\
\sigma_0^2 & (\tau+1)\sigma_0^2 \end{array} \right) \right),\label{prior}
\end{equation}
where $\pmb{\delta}_0$ is the bivariate unit point mass distribution at the origin, and $N_2(\pmb{\eta},\Sigma)$ denotes the bivariate Gaussian distribution with expectation $\pmb{\eta}$ and covariance matrix $\Sigma$.

Since the observed effect size $\widehat{\mu}^{(j)}$ asymptotically follows Gaussian distribution $N(\mu^{(j)}, (\sigma^{(j)})^2)$, the test statistics vector $\pmb{Z}_i=(Z^{(1)}_i,Z^{(2)}_i)^T$ with prior (\ref{prior}) follows two-component Gaussian mixture distribution:
\begin{equation}
\pmb{Z}_i \sim \pi_0 N_2(\mathbf{0}, I)+(1-\pi_0) N_2(\mathbf{0}, I+\Sigma_i), \text{ where } \Sigma_i=\left( \begin{array}{cc}
(\tau+1)(\frac{\sigma_0}{\sigma^{(1)}_i})^2 & \frac{\sigma_0^2}{\sigma^{(1)}_i\sigma^{(2)}_i} \\
\frac{\sigma_0^2}{\sigma^{(1)}_i \sigma^{(2)}_i} & (\tau+1)(\frac{\sigma_0}{\sigma^{(2)}_i})^2 \end{array} \right).
\end{equation}
Here, $I$ is the identity matrix.

In order to obtain the global behavior of all SNPs, we need to obtain the marginal distribution of the test statistic vectors of all SNPs. Overall, the test statistic vector $\pmb{Z}$ follows
\begin{equation}
\pmb{Z}\sim \pi_0 N_2(\mathbf{0}, I)+ \frac{1-\pi_0}{m_1} \sum_{i\in S_1} N_2(\mathbf{0}, I+\Sigma_i). \label{gmm0}
\end{equation}
Here $S_1$ is the index set of all associated SNPs, and $m_1$ is the corresponding cardinality of $S_1$. $S_1$ is normally unknown.

The above Gaussian mixture model is computationally difficult due to the large number of components ($m_1$ is normally in the range of hundred to thousand). To simplify the model, we use a $K$-component Gaussian mixture model to approximate the non-null components, i.e.,
\begin{equation}
\frac{1-\pi_0}{m} \sum_{i=1}^m N_2(\mathbf{0}, I+\Sigma_i)\approx \sum_{k=1}^K \pi_{1k} N_2(\mathbf{0}, I+\bar{\Sigma}_k), \text{ where } \sum_{k=1}^K \pi_{1k}=1-\pi_0.
\end{equation}
Then we reduce the distribution of $\pmb{Z}$ to a $(K+1)$-component Gaussian mixture model:
\begin{equation}
\pmb{Z}\sim \pi_0 N_2(\mathbf{0}, I)+\sum_{k=1}^K \pi_{1k} N_2(\mathbf{0}, I+\bar{\Sigma}_k).\label{gmm}
\end{equation}

There are some unknown parameters $\pmb{\pi}_1=(\pi_{11},\dots,\pi_{1K})^T$ and $\pmb{\bar{\Sigma}}=\{\bar{\Sigma}_1,\dots,\bar{\Sigma}_K\}$ in the above mixture model. \cite{dempster1977maximum} proposed an EM-algorithm to estimate parameters with unobserved latent variables. With the observed vectors of summary statistics $\pmb{z}_i$ ($i=1,\dots,m$), we use the EM-algorithm to estimate the parameters $\pmb{\pi}_1$ and $\pmb{\bar{\Sigma}}$ in the Gaussian mixture model (\ref{gmm}). Please note that $\pi_0$ is always much larger than any entry of $\pmb{\pi}_1$ in the GWAS setting. Hence, a Dirichlet$(\beta_0,\pmb{0}^T)$ prior is added for the proportions $(\pi_0,\pmb{\pi_1}^T)$. This is the same penalty strategy proposed by \cite{muralidharan2010empirical}. Our experiments show that the rejection regions are not sensitive to the penalization parameter $\beta_0$ and the number of mixture components in the associated SNPs $K$. In our default setting, $\beta_0=m/5$ and $K=2$.

Denote the probability density function (pdf) of bivariate normal distribution $N_2(\mathbf{0},I)$ as $f_0(x_1,x_2)$ and the pdf of $N_2(\mathbf{0}, I+\bar{\Sigma}_k)$ as $f_1(x_1,x_2|\bar{\Sigma}_k)$. The Jlfdr reads
\begin{equation}
\text{Jlfdr}(\mathbf{z})=\frac{\pi_0 f_0(z_1, z_2)}{\pi_0 f_0(z_1, z_2)+\sum_{k=1}^K \pi_{1k} f_1(z_1, z_2|\bar{\Sigma}_k)}\label{Jlfdr}.
\end{equation}
After calculating the Jlfdr, we approximate Fdr as
\begin{equation}
\text{Fdr}(\mathcal{R})=E(\text{Jlfdr}(\mathbf{z})\big| \mathbf{z}\in \mathcal{R})\approx \frac{1}{|\{\mathbf{z}\in \mathcal{R}\}|} \sum_{\mathbf{z}\in \mathcal{R}} \text{Jlfdr}(\mathbf{z}).
\end{equation}

We determine the optimal rejection region $\mathcal{R_O}$ by Jlfdr-thresholding, which determines the rejection region with $\text{Jlfdr}(\mathbf{z})$ smaller than the threshold $t(q)$. To determine the threshold $t(q)$, we sort the calculated Jlfdr values of each SNP in an ascending order first. Denote the $a$-th Jlfdr value as $\text{Jlfdr}_a$. We can approximate the Fdr of the region $\mathcal{R}_a=\{\mathbf{z} \big| \text{Jlfdr}(\mathbf{z})\leq \text{Jlfdr}_a\}$ as
\begin{equation}
\text{Fdr}(\mathcal{R}_a)\approx \frac{1}{a} \sum_{b=1}^{a} \text{Jlfdr}_{b}.\label{Fdr}
\end{equation}
We use $c$ to denote the largest $a$ such that $\text{Fdr}(\mathcal{R}_a)\leq q$, namely
\begin{equation}
c=\max \{a\big| \text{Fdr}(\mathcal{R}_a)\leq q\}.
\end{equation}
Then the Jlfdr threshold $t(q)$ is $\text{Jlfdr}_{c}$. We reject all SNPs with $\text{Jlfdr}(\pmb{z})\leq t(q)$.

We present the detailed steps of the Jlfdr-based method in Algorithm \ref{algorithm}.

\begin{algorithm}[!htbp]
 \caption{Jlfdr-based method for summary-statistics-based joint analysis}\label{algorithm}
 \centering
 \begin{algorithmic}
 \BState \emph{Inference using the EM-algorithm:}
 \State \textbf{Initialize} $\pmb{\pi}_1$ and $\pmb{\bar{\Sigma}}$
 \Repeat
    \State \textbf{E Step:}
    \[\pi_0^t \gets 1-\sum_{k=1}^K \pi_{1k}^t\]
    \[h_{i0} \gets \frac{\pi_0^t f_0(z^{(1)}_i, z^{(2)}_i)}{\pi_0^t f_0(z^{(1)}_i, z^{(2)}_i)+\sum_{k=1}^K \pi_{1k} f_1(z^{(1)}_i, z^{(2)}_i|\Sigma_{1k}^t)},\ i=1,\cdots,m \]
    \[h_{il}\gets \frac{\pi_{1l}^t f_1(z^{(1)}_i, z^{(2)}_i|\Sigma_{1l}^t)}{\pi_0^t f_0(z^{(1)}_i, z^{(2)}_i)+\sum_{k=1}^K \pi_{1k} f_1(z^{(1)}_i, z^{(2)}_i|\Sigma_{1k}^t)},\ l=1,\cdots,K \]
    \State \textbf{M Step:}
     \[\pi_{1l}^{t+1} \gets  \frac{\sum_{i=1}^m h_{il}}{m+\beta_0} \]
     \[\Sigma_{1l}^{t+1} \gets  \frac{\sum_{i=1}^m h_{il}\pmb{z}_i\pmb{z}_i^T}{\sum_{i=1}^m h_{il}}-I, l=1,\cdots,K\nonumber\]
 \Until{$\pmb{\pi}_1$ and $\pmb{\bar{\Sigma}}$ converge}
 \BState \emph{Jlfdr-thresholding:}
 \State \textbf{Initialize} $t(q) \gets 0$
 \State Calculate Jlfdr for each SNP using Eq. (\ref{Jlfdr}) with inferred $\pmb{\pi}_1$ and $\pmb{\bar{\Sigma}}$.
 \State Sort calculated Jlfdr in ascending order
 \For{$a \gets 1$ to $m$}
   \State Calculate $\text{Fdr}(\mathcal{R}_a)$ using Eq. (\ref{Fdr}).
   \If{$\text{Fdr}(\mathcal{R}_a) > q$,}
       \State $t(q) \gets \text{Jlfdr}_{a-1}$; break
   \EndIf
 \EndFor
\State Output: the SNPs with $\text{Jlfdr}\leq t(q)$
 \end{algorithmic}
\end{algorithm}

\subsection{Relationship between Jlfdr-based method and meta-analysis methods}\label{relation}
We have the following theorem about the rejection region of Jlfdr-based method when using the Gaussian mixture model:

\begin{them}\label{them2}
In the Gaussian mixture model (\ref{gmm0}), the rejection region of the Jlfdr-based method is
\begin{equation}
\mathcal{R}_1=\{\mathbf{z}\big| \sum_{i\in S_1} exp(\mathbf{z}^T(I-(I+\Sigma_i)^{-1})\mathbf{z}) \geq C_1\},
\end{equation}
where $C_1$ is a constant determined by $\text{Fdr}(\mathcal{R_O}_1)=q$. If no heterogeneity exists between studies, the rejection region is
\begin{equation}
\mathcal{R}_2=\{\mathbf{z}\big| |\pmb{\alpha}^T \pmb{z}|\geq C_2\},
\end{equation}
where $\pmb{\alpha}=(\sqrt{n^{(1)}},\sqrt{n^{(2)}})^T$, and $C_2$ is a constant determined by $\text{Fdr}(\mathcal{R_O}_2)=q$.
\end{them}

We present the proof of the above theorem in the Supplementary Note.

Meta-analysis methods are the most commonly used summary-statistics-based joint analysis methods. In meta-analysis, we usually calculate the weighted average of effect sizes in different studies. Dividing the weighted average effect size by its standard error yields a new $z$-value-based test statistic. There are two kinds of models in the meta-analysis: fixed-effects model and random-effects model.

In the fixed-effects model, we assume that the underlying true effect sizes in different studies are identical. This corresponds to $\tau=0$ in the Gaussian mixture model \ref{gmm0}. The optimal weighting strategy is the inverse-variance weighting since it minimizes the variance of the weighted average. Each effect size is weighted by the inverse of its variance, i.e.,
\begin{equation}\label{metaMu}
\widehat{\mu}_w=\frac{w_1 \widehat{\mu}^{(1)}+w_2 \widehat{\mu}^{(2)}}{w_1+w_2}, \text{ with } w_j=\frac{1}{(\sigma^{(j)})^2}, j=1,2.
\end{equation}
Here $\widehat{\mu}_{w}$ is the weighted average effect size. Its standard error is
\begin{equation}\label{metaSE}
\sigma_w=\sqrt{\frac{1}{w_1+w_2}}.
\end{equation}
Dividing $\widehat{\mu}_w$ by $\sigma_w$ yields the new test statistic $z_w$. We have the following theorem about the rejection region of the fixed-effects meta-analysis method (See the Supplementary Note for detailed proof):

\begin{them}\label{them3}
The rejection region of the fixed-effects meta-analysis method is asymptotically
\begin{equation}
\mathcal{R}_3=\{\mathbf{z}\big| |\pmb{\alpha}^T \pmb{z}|\geq C_3\},
\end{equation}
where $\pmb{\alpha}=(\sqrt{n^{(1)}},\sqrt{n^{(2)}})^T$, and $C_3$ is a constant determined by $\text{Fdr}(\mathcal{R}_3)=q$. .
\end{them}

This kind of rejection region is illustrated in Figure \ref{Figure1}. The region coincides with the rejection region of the Jlfdr-based method when no heterogeneity exists between studies. Hence, the Jlfdr-based method and the fixed-effects meta-analysis method will have the same performance. In contrast, if heterogeneity exists between studies, the rejection regions determined by the Jlfdr-based method and the fixed-effects meta-analysis method are different. According to Theorem \ref{them1}, the rejection region determined by the Jlfdr-based method can achieve the highest power among all summary-statistics-based joint analysis methods when controlling the Fdr. In other words, the Jlfdr-based method is more powerful than the fixed-effects meta-analysis method.

In the random-effects model, we assume that the true effect sizes in different studies are not identical and follow a distribution. Then we adjust the weights by incorporating the between-study variance. The weighted average effect size is
\begin{gather}
\widehat{\mu}_{w}^*=\frac{w_1^* \widehat{\mu}^{(1)}+w_2^* \widehat{\mu}^{(2)}}{w_1^*+w_2^*},\nonumber\\
\text{ with } w_j^*=\frac{1}{(\sigma^{(j)})^2+\hat{\Delta}^2}, j=1,2,\text{ and }  \hat{\Delta}^2=\max \left( 0, \frac{w_1(\widehat{\mu}^{(1)}-\widehat{\mu}_w)^2+w_2(\widehat{\mu}^{(2)}-\widehat{\mu}_w)^2-1}{(w_1+w_2)-(w_1^2+w_2^2)/(w_1+w_2)} \right).
\end{gather}
Its standard error is
\begin{equation}
\sigma_w^*=\sqrt{\frac{1}{w_1^*+w_2^*}}.
\end{equation}
Dividing $\widehat{\mu}_w^*$ by $\sigma_w^*$ yields the new test statistic $z_w^*$. We have the following theorem about the rejection region of the random-effects meta-analysis method (See the Supplementary Note for detailed proof):
\begin{them}\label{them4}
The rejection region of the random-effects meta-analysis method is asymptotically
\begin{gather}
\mathcal{R}_4=\{\mathbf{z}\big| \left|\frac{\pmb{\alpha}^T V \pmb{z}}{\pmb{\alpha}^T V \pmb{\alpha}}\right|\geq C_4\} \nonumber\\
\text{ with } V=\left( \begin{array}{cc}
1/(1+\alpha_1^2 s) & 0 \\
0 & 1/(1+\alpha_2^2 s) \end{array} \right) \text{ and }  s=\max \left(0, \frac{||\pmb{\alpha}||_2^2 \left[||\pmb{z}||_2^2-(\pmb{\alpha}^T \pmb{z}/||\pmb{\alpha}||_2)^2-1 \right]}{(\pmb{\alpha}^T\pmb{\alpha})^2-||\pmb{\alpha}^{o 2}||_2^2}\right).
\end{gather}
Here $\pmb{\alpha}=(\sqrt{n^{(1)}},\sqrt{n^{(2)}})^T$, $\pmb{\alpha}^{o 2}=(n^{(1)},n^{(2)})^T$ which is the 2nd Hadamard power of $\pmb{\alpha}$, and $C_4$ is a constant determined by $\text{Fdr}(\mathcal{R}_4)=q$.
\end{them}

If no heterogeneity exists between studies, the random-effects meta-analysis method is less powerful than the fixed-effects meta-analysis method and the Jlfdr-based method. If heterogeneity exists between studies, we usually need a large number of studies to estimate the between-study variance precisely in the random-effects meta-analysis. Since we usually only have a few GWASs of the same diseases in the same population, the random-effects meta-analysis is not powerful enough. The Jlfdr-based method overcomes this problem by borrowing information from all genotyped SNPs. In any case, the rejection region determined by the Jlfdr-based method and the random-effects meta-analysis method are different. According to Theorem \ref{them1}, the Jlfdr-based method is more powerful than the random-effects meta-analysis method.

\section{Results}\label{results}

\subsection{Simulation experiments}
We use simulation experiments to demonstrate that the Jlfdr-based method is more powerful than the commonly used meta-analysis methods in analyzing summary statistics from multiple GWASs.

In our simulation experiments, we fix the sample size at 10000 in study 1. We conduct experiments with different sample sizes of 5000, 10000 and 15000 in study 2. The sample size ratios $n^{(2)}/n^{(1)}$ are 0.5, 1 and 1.5 correspondingly. The individual numbers in the control group and case group are the same in both studies, and the number of SNPs is $m=1\times 10^6$. We simulate the minor allele frequency of each SNP according to uniform distribution $U(0.05, 0.5)$. The proportion of the associated SNPs is $5\%$. For associated SNPs, the expected $log$-odds ratio $\mu^{(j)}$ in each study is simulated according to the following model:
\begin{eqnarray}
\mu^{(j)}\big| \mathcal{H}_1 &\sim& N(\mu, \tau \sigma_0^2) \nonumber \\
\mu \big| \mathcal{H}_1 &\sim& N(0,\sigma_0^2),
\end{eqnarray}
where $\sigma_0^2=0.04$. In the homogeneous setting, $\tau=0$. In the heterogeneous setting, $\tau=0.5$. For non-associated SNPs, the expected $log$-odds ratio $\mu^{(j)}$ is $0$. The prevalence of the disease is $1\%$. We use the $log$-odds ratio test to detect associations in our experiments.

We use the Jlfdr-based method, the fixed-effects meta-analysis method and the random-effects meta-analysis method to jointly analyze summary statistics from study 1 and study 2. The Fdr is controlled at $q=5\times 10^{-5}$. In the fixed-effects meta-analysis and the random-effects meta-analysis, we use the one-dimensional mixture method \cite{muralidharan2010empirical} to control the Fdr at $q$.

\begin{figure}[!htbp]
    \centering
    \includegraphics[width=0.95\textwidth]{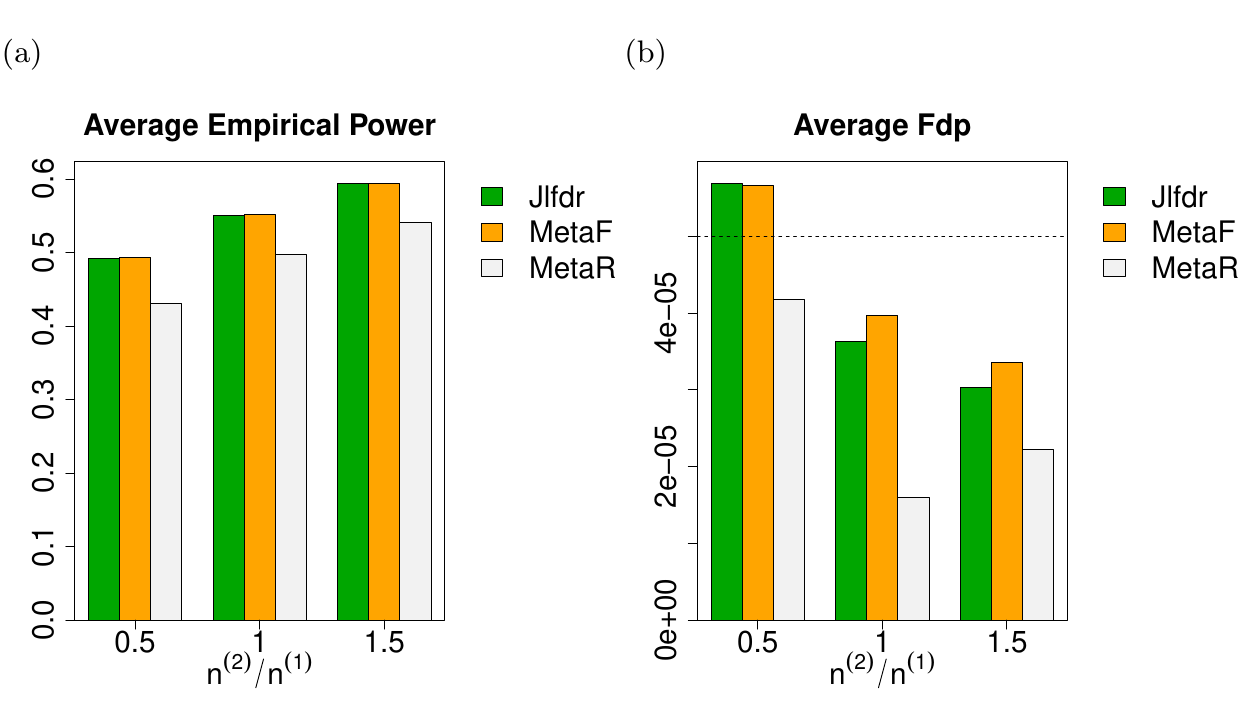}
    \caption{\textbf{(a) The average empirical power and (b) the average Fdp in the homogeneous setting ($\tau=0$) of the simulation experiment.} The experiments are repeated 10 times with different sample size ratios ($n^{(2)}/n^{(1)}=0.5$, $1$ and $1.5$). The average Fdp of the three methods (the Jlfdr-based method (Jlfdr), the fixed-effects meta-analysis method (MetaF) and the random-effects meta-analysis method (MetaR)) are about $5\times 10^{-5}$. When controlling Fdr at the same level, the Jlfdr-based method and the fixed-effects meta-analysis method have almost the same average empirical power. The subtle differences are due to random initial choices of the EM-algorithm and the Fdr approximations used in Eq. (\ref{Fdr}). }\label{Figure2}
\end{figure}

In the homogeneous setting ($\tau=0$), each SNP shares the same expected effect size between the two studies. Figure \ref{Figure2} presents the average empirical power and the average Fdp of 10 experimental runs using different methods. The average Fdp is well controlled in all methods. In this setting, the average empirical powers are almost the same in the Jlfdr-based method and the fixed-effects meta-analysis method. The subtle differences are due to random initial choices of the EM-algorithm and the Fdr approximations used in Eq. (\ref{Fdr}). This verifies the previous statement about the equivalence between the Jlfdr-based method and the fixed-effects meta-analysis method in the homogeneous setting.

\begin{figure}[!htbp]
    \centering
    \includegraphics[width=0.8\textwidth]{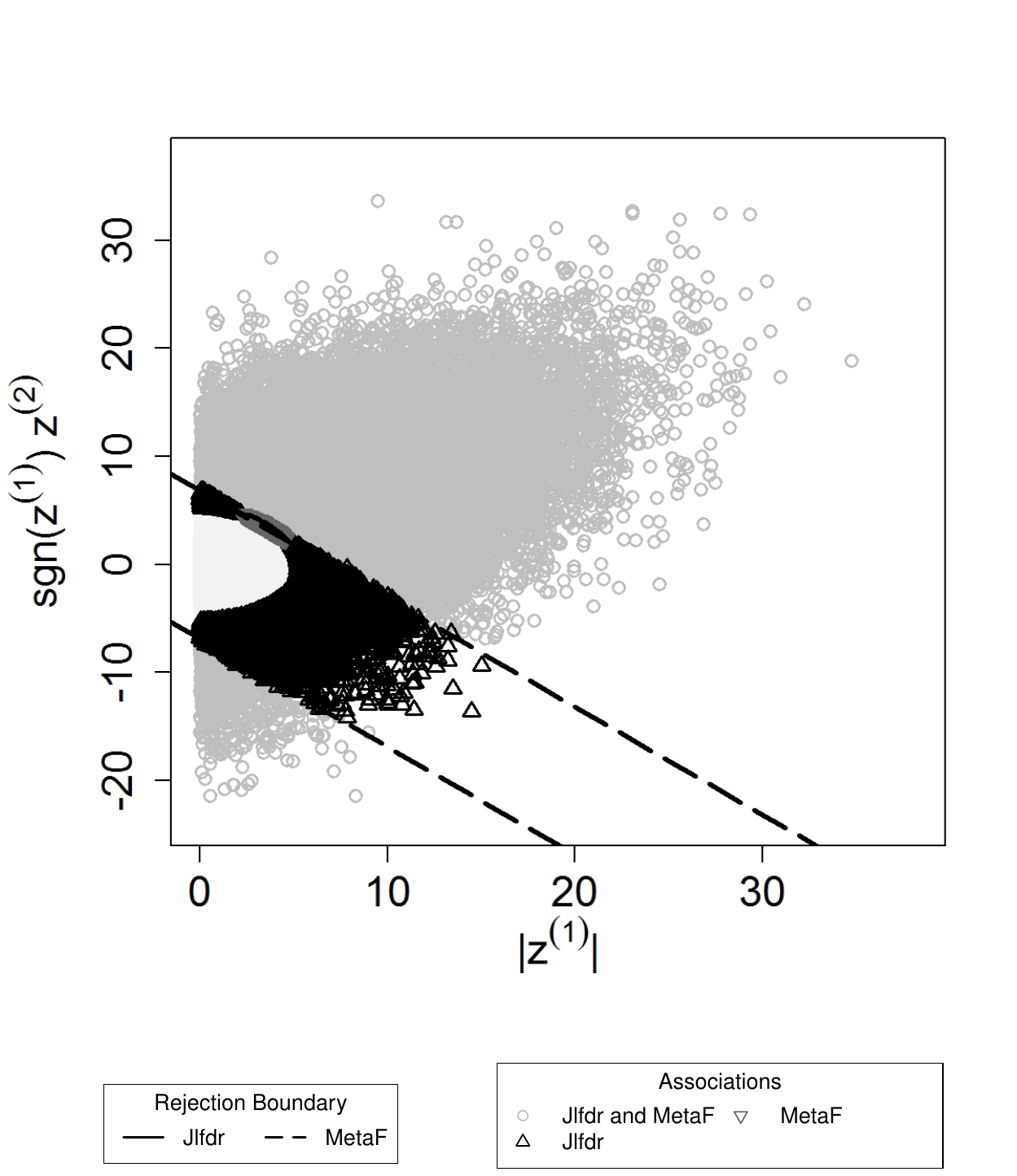}
    \caption{\textbf{The discovered associations in the heterogeneous setting ($\tau=0.5$) of the simulation experiment.} Both the first and second studies have $10000$ individuals. For each SNP, the pair of summary statistics $(z^{(1)}, z^{(2)})$ is plotted with transformation $(|z^{(1)}|, sgn(z^{(1)})z^{(2)})$. We use light grey circles to represent the associations discovered by both the Jlfdr-based method and fixed-effects meta-analysis method. We use black upward-pointing triangles and dark grey downward-pointing triangles to represent the associations only discovered by the Jlfdr-based method and the fixed-effects meta-analysis method, respectively. The rejection boundary in the Jlfdr-based method is plotted as the solid curve. The rejection boundary in the fixed-effects meta-analysis method is plotted as the dashed straight line. The Jlfdr-based method discovered more associations overall than the meta-analysis method, although it also misses some associations identified by the meta-analysis method.}\label{Figure3}
\end{figure}

\begin{figure}[!htbp]
    \centering
    \includegraphics[width=0.95\textwidth]{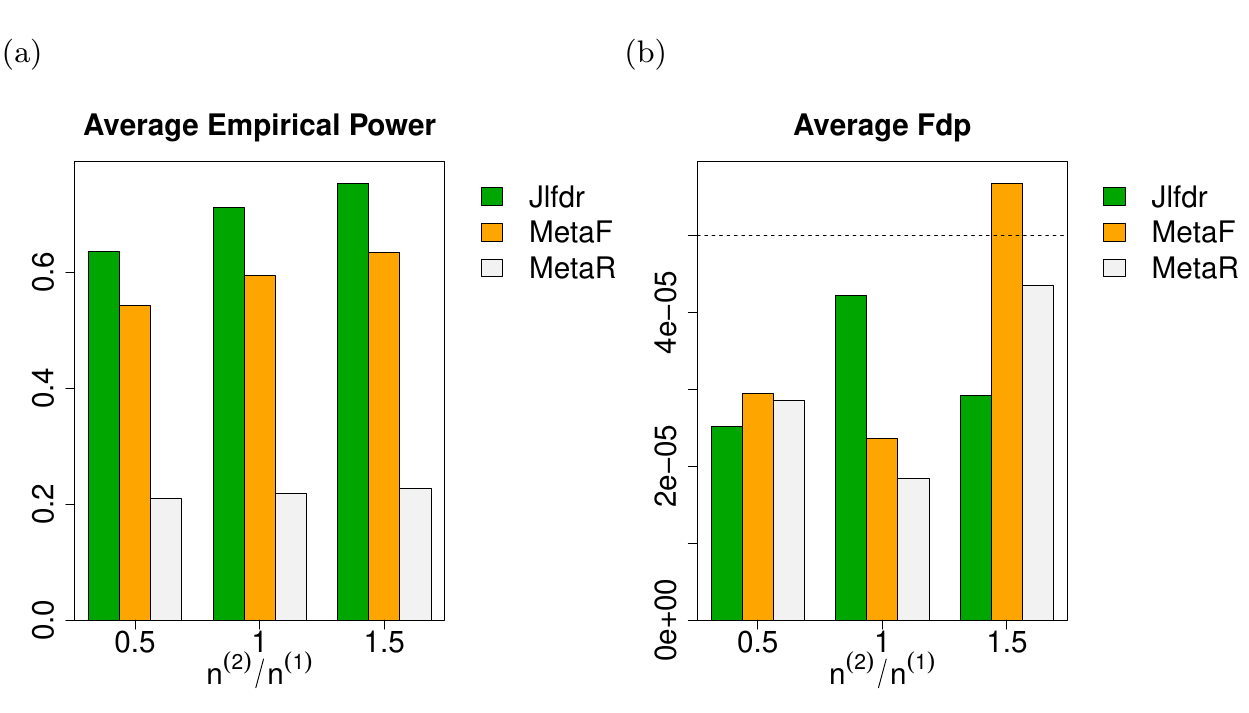}
    \caption{\textbf{(a) The average empirical power and (b) the average Fdp in the heterogeneous setting ($\tau=0.5$) of the simulation experiment.} We ran experiments 20 times with different sample size ratios ($n^{(2)}/n^{(1)}=0.5$, $1$ and $1.5$). The average Fdp values in three methods are about $5\times 10^{-5}$. When controlling Fdr at the same level, our proposed Jlfdr-based method can achieve higher power than the other methods in every sample size ratio setting.}\label{Figure4}
\end{figure}

In the heterogeneous setting ($\tau=0.5$), the expected effect sizes of each SNP vary between studies. Figure \ref{Figure3} plots the discovered associations using the Jlfdr-based method and the fixed-effects meta-analysis method in one run when $n^{(2)}=10000$. Although the Jlfdr-based method missed some associations detected by the fixed-effects meta-analysis method, it identifies more associations than the meta-analysis method. We ran the simulation experiments 10 times for the sample size ratio $n^{(2)}/n^{(1)}=0.5, 1$ and $1.5$. Figure \ref{Figure4} shows the average empirical power and the average Fdp. The average Fdp using all three methods are about $q=5\times 10^{-5}$ in all sample size ratio settings. From the figure, we can see that the Jlfdr-based method can achieve higher power than the other methods when controlling Fdr at the same threshold.

\subsection{Real data applications}
\subsubsection{SCZ data from PGC}
We jointly analyze the summary statistics from schizophrenia (SCZ) studies conducted by the Psychiatric Genomics Consortium (PGC). The summary statistics from two SCZ studies, Sweden+SCZ1 \citep{ripke2013genome} and SCZ2 \citep{schizophrenia2014biological}, are available from the PGC. Sweden+SCZ1 is a large-scale meta-analysis of Swedish and mixed-European ancestry individuals that comprises 13,833 schizophrenia cases and 18,310 controls in the analysis. We use it as Study 1. SCZ2 is a larger-scale meta-analysis that comprises 36,989 schizophrenia cases and 113,075 controls. The analysis includes the individuals which have been analyzed in Sweden+SCZ1. By using the following inverse meta-analysis formula, we obtain the summary statistics from the meta-analysis comprising the individuals only be analyzed in SCZ2. The formula is
\begin{equation}
z^{(2)}=\frac{z_w/\sigma_w-z^{(1)}/\sigma^{(1)}}{1/(\sigma_w)^2-1/(\sigma^{(1)})^2}.
\end{equation}
We use $z^{(2)}$ as the summary statistics of Study 2. We remove the SNPs with $p$-value$<0.01$ in the test of homogeneity. After that, there are $m=8,157,410$ SNPs remaining.

We use the Jlfdr-based method, the fixed-effects meta-analysis method and the random-effects meta-analysis method to jointly analyze the summary statistics from two studies. The Fdr is controlled at $q=5\times 10^{-5}$. We adopt the one-dimensional mixture method to control the Fdr at $q$ in meta-analysis methods.

\begin{figure}[!htbp]
    \centering
    \includegraphics[width=0.9\textwidth]{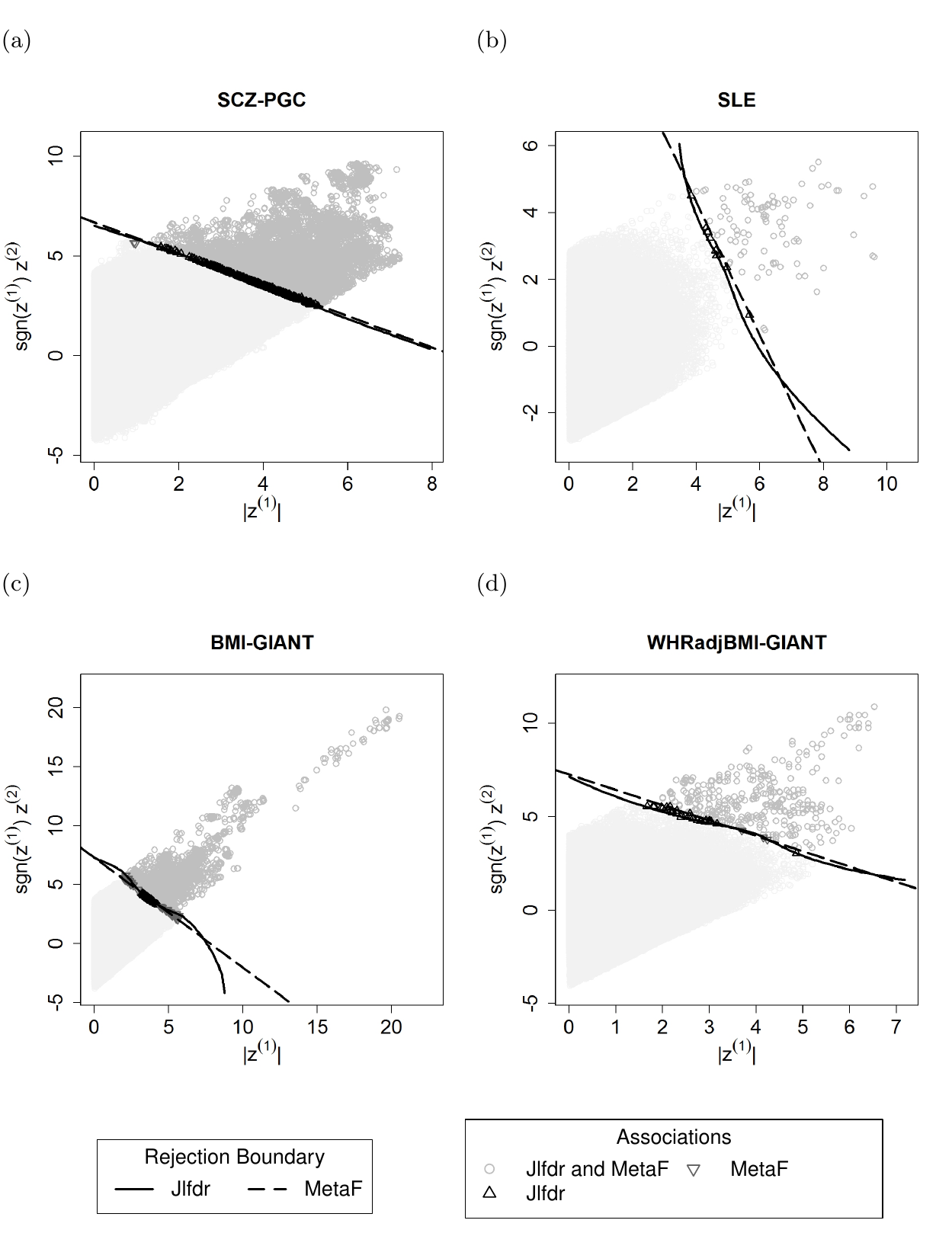}
    \caption{\textbf{The rejection region determined in the empirical datasets: }(a) SCZ data from the PGC; (b) SLE data from dbGaP; (c) BMI data from the GIANT; (d) WHRadjBMI data from the GIANT. The descriptions of the three datasets are presented in the main text. For each SNP, the vector of summary statistics $(z^{(1)}, z^{(2)})$ is plotted with transformation $(|z^{(1)}|, sgn(z^{(1)})z^{(2)})$. We use light grey circles to represent the associations discovered by both the Jlfdr-based method and the fixed-effects meta-analysis method. We use black upward-pointing triangles and dark grey downward-pointing triangles to represent the associations only discovered by the Jlfdr-based method and the fixed-effects meta-analysis method, respectively. }\label{Figure5}
\end{figure}

Figure \ref{Figure5}(a) plots the discovered associations using the Jlfdr-based method and the fixed-effects meta-analysis method. The Jlfdr-based method identifies more associations. Table \ref{SCZtable} shows the numbers of discovered associations and the rejection criteria of the different analysis methods. Besides the loci discovered by meta-analysis methods, there are eight novel loci discovered by the Jlfdr-based method. Each locus is separated by at least 500 kilobases (kb) or a weak linkage disequilibrium ($r^2<0.1$). The SNPs showing the most significant association with SCZ in these novel loci are presented in Supplementary Table 1.

\begin{table}[!htbp]
\centering
\caption{\textbf{The rejection criterion and the number of identified associations in SCZ data from the PGC.} $z_{MF}$ and $z_{MR}$ are the combined $z$-values in the fixed-effects meta-analysis and random-effects meta-analysis, respectively.}\label{SCZtable}
\begin{tabular}{c|c|c}
Method & Rejection Criterion & \#\{Identified SNPs\}\\ \hline
Jlfdr-based method & $\text{Jlfdr}(\pmb{z})\leq 3.206\times 10^{-4}$ & 13405\\
Fixed-effects meta-analysis &  $|z_{MF}|\geq 5.273$ & 13014\\
Random-effects meta-analysis & $|z_{MR}|\geq 5.352$ & 8348\\\hline
\end{tabular}
\end{table}

\subsubsection{SLE data from dbGaP}
We conduct summary-statistics-based joint analysis in systemic lupus erythematosus (SLE) data from phs000122.v1.p1 and phs000216.v1.p1 in dbGaP \citep{mailman2007ncbi, tryka2014ncbi}. We use the study phs000122.v1.p1, in which there are 1,311 SLE cases and 3,340 controls, as Study 1, and we use the study phs000216.v1.p1, in which there are 706 cases and 353 controls, as Study 2. The individuals in the first study are all North Americans of European descent, and those in the second study are all females of European ancestry. We use the following quality control procedures for both studies:
\begin{enumerate}
\item Missing data control: The SNPs with a missing data rate larger than $1\%$ are discarded.
\item Minor allele frequency control: The SNPs with minor allele frequency less than $0.05$ in either case group or control group are discarded.
\item Hardy-Weinberg equilibrium control: In the Hardy-Weinberg equilibrium test, the SNPs with $p$-values less than $0.001$ in either case group or control group are discarded.
\item Homogeneity control: In the homogeneity test, SNPs with $p$-values less than $0.01$ are discarded.
\end{enumerate}
After the quality control steps, there are $m=195,318$ autosome SNPs remaining.

We use $q=5\times 10^{-5}$ as the Fdr threshold in all analyses. \ref{Figure5}(b) plots the associations discovered by the Jlfdr-based method and the fixed-effects meta-analysis method. The Jlfdr-based method discovers more associations than the meta-analysis methods. Table \ref{SLEtable} lists the numbers of the associations discovered using the different methods. Besides the loci discovered by meta-analysis methods, there are three novel loci discovered by the Jlfdr-based method. The loci are separated by at least 500kb or a weak linkage disequilibrium ($r^2<0.1$). The most significant associations in these novel loci can be seen in Supplementary Table 2.

\begin{table}[!htbp]
\centering
\caption{\textbf{The rejection criterion and the number of identified associations in SLE data from dbGaP.} $z_{MF}$ and $z_{MR}$ are the combined $z$-values in the fixed-effects meta-analysis and random-effects meta-analysis, respectively.}\label{SLEtable}
\begin{tabular}{c|c|c}
Method & Rejection Criterion & \#\{Identified SNPs\}\\ \hline
Jlfdr-based method & $\text{Jlfdr}(\pmb{z})\leq 5.543\times 10^{-4}$ & 106\\
Fixed-effects meta-analysis &  $|z_{MF}|\geq 5.508$ & 94\\
Random-effects meta-analysis & $|z_{MR}|\geq 5.586$ & 54 \\\hline
\end{tabular}
\end{table}

\subsubsection{BMI data from GIANT}
We jointly analyze summary statistics from body mass index (BMI) studies conducted by the Genetic Investigation of ANthropometric Traits (GIANT) consortium \citep{locke2015genetic}. We use the joint GWAS and metabochip meta-analysis of 152,893 European men as Study 1, and we use the joint GWAS and metabochip meta-analysis of 171,977 European women as Study 2. There are $m=2,466,338$ autosome SNPs passing the homogeneity control ($p$-value$\geq 0.01$).

We use $q=5\times 10^{-5}$ as the Fdr threshold in all analyses. Figure \ref{Figure5}(c) plots the associations discovered by the Jlfdr-based method and the fixed-effects meta-analysis method. The Jlfdr-based method discovers more associations than meta-analysis methods. Table \ref{BMItable} shows the number of discovered associations and the corresponding rejection criterion of each method. There are six novel loci discovered by the Jlfdr-based method. The SNPs showing the most significant associations in these novel loci are listed in Supplementary Table 3.

\begin{table}[!htbp]
\centering
\caption{\textbf{The rejection criterion and the number of identified associations in BMI data from the GIANT.} $z_{MF}$ and $z_{MR}$ are the combined $z$-values in the fixed-effects meta-analysis and random-effects meta-analysis, respectively.}\label{BMItable}
\begin{tabular}{c|c|c}
Method & Rejection Criterion & \#\{Identified SNPs\}\\ \hline
Jlfdr-based method & $\text{Jlfdr}(\pmb{z})\leq 3.722\times 10^{-4}$ & 2717\\
Fixed-effects meta-analysis &  $|z_{MF}|\geq 5.336$ & 2667\\
Random-effects meta-analysis & $|z_{MR}|\geq 5.383$ & 2186 \\\hline
\end{tabular}
\end{table}

\subsubsection{WHRadjBMI data from GIANT}
We conduct joint analysis in waist-to-hip ratio after adjusting for BMI (WHRadjBMI) studies from GIANT consortium \citep{shungin2015new}. We use the joint GWAS and metabochip meta-analysis of 93,480 European men as Study 1, and we use the joint GWAS and metabochip meta-analysis of 116,742 European women as Study 2. There are $m=2,127,324$ autosome SNPs passing the homogeneity control ($p$-value$\geq 0.01$).

Figure \ref{Figure5}(d) highlights the associations discovered by the Jlfdr-based method and the fixed-effects meta-analysis method. The Jlfdr-based method identifies more associations than meta-analysis methods when controlling Fdr at the same level $q=5\times 10^{-5}$. Table \ref{WHRadjBMItable} shows the number of the discovered associations and the corresponding rejection criterion of each method. Besides the loci discovered by meta-analysis methods, there are four novel loci discovered by the Jlfdr-based method. The details of the most significant SNPs in these loci are listed in Supplementary Table 4.

\begin{table}[!htbp]
\centering
\caption{\textbf{The rejection criterion and the number of identified associations in WHRadjBMI data from the GIANT.} $z_{MF}$ and $z_{MR}$ are the combined $z$-values in the fixed-effects meta-analysis and random-effects meta-analysis, respectively.}\label{WHRadjBMItable}
\begin{tabular}{c|c|c}
Method & Rejection Criterion & \#\{Identified SNPs\}\\ \hline
Jlfdr-based method & $\text{Jlfdr}(\pmb{z})\leq 5.750\times 10^{-4}$ & 452\\
Fixed-effects meta-analysis &  $|z_{MF}|\geq 5.617$ & 420\\
Random-effects meta-analysis & $|z_{MR}|\geq 5.742$ & 192 \\\hline
\end{tabular}
\end{table}

\section{Discussion}\label{discussion}

%


Both the Jlfdr-based method and the meta-analysis methods jointly analyze summary statistics from multiple GWASs. Meta-analysis methods collapse the test statistics of all studies into a weighted average value for each SNP, which is simpler than the Jlfdr-based method. When no heterogeneity exists between studies, the Jlfdr-based method will degenerate to the fixed-effects meta-analysis method. This can be understood by the fact that there is no information loss during the collapsing when all studies are homogeneous. When heterogeneity exists between studies, however, the Jlfdr-based method can achieve higher power than the fixed-effects meta-analysis method. This is understandable as information about heterogeneity is lost during collapse when using the meta-analysis method. Since heterogeneity widely exists in most cases, we suggest to use the Jlfdr-based method instead of meta-analysis methods to jointly analyze summary statistics from multiple GWASs.

This paper proves that the Jlfdr-based method is the most powerful summary-statistics-based joint analysis method when the underlying distribution of the test statistics is known. In reality, we only know the theoretical distribution under a null hypothesis. The distribution under alternative hypotheses is usually unknown. Hence, in the implementation of the Jlfdr-based method, we assume test statistics follow the Gaussian mixture model. Then we use the EM-algorithm to infer parameters in the mixture model. Violation of the model assumptions and inaccuracy of parameters estimation will decrease the performance of the Jlfdr-based method.



We assume an independence between SNPs in the Gaussian mixture model. However, correlations between nearby SNPs often exist, which is known as linkage disequilibrium. We may further improve the Jlfdr-based method by taking advantage of the dependency information between SNPs.

\section{Conclusion}\label{conclusion}
Jointly analyzing data sets from multiple GWASs is a common strategy to discover associations. Since it is usually difficult to access individual-level genotyping data, summary-statistics-based joint analysis has become popular for jointly analyzing data sets from multiple GWASs. Among different summary-statistics-based joint analysis methods, we prefer the method with higher Bayesian power when Fdr is controlled at the same level, because it can discover more associations. With this criterion, we propose the Jlfdr-based method. It is the most powerful summary-statistics-based method. Simulation and empirical experiments demonstrate its superior performance over traditional meta-analysis methods.

\section*{Acknowledgement}
This paper was partially supported by a grant under the Theme-based Research Scheme (project T12-402/13N) of the Hong Kong Research Grant Council (RGC).

\bibliography{Jlfdr}

\end{document}